\documentclass[superscriptaddress,aps,preprintnumbers,amsmath,showpacs,amssymb,prd,nofootinbib,reprint]{revtex4-1}
\usepackage{bm, color} 
 \usepackage{amssymb,amsfonts,slashed,amsthm,amsmath,graphicx, soul}

\usepackage{footmisc}
\usepackage{hyperref}
\begin{document}

\renewcommand{\thesection}{\Alph{section}}

\newcommand{\mplank}{\textrm{M}_{\textrm{P}}}
\newcommand{\mg}{m_{\gamma^{\prime}}}
\newcommand{\higgs}{H_{\scriptsize \rm h}}
\newcommand{\higgst}{\tilde{H}_{\scriptsize \rm h}}
\renewcommand{\Re}{\mathrm{Re}}
\newcommand{\MF}{{\sf B}}

\newcommand{\muu}{m_{\gamma^{\prime}}}
\newcommand{\chie}{\chi_{\rm eff}}
\newcommand{\ve}[2]{\left(
\begin{array}{c}
 #1\\
#2
\end{array}
\right)
}

\renewcommand\({\left(}
\renewcommand\){\right)}
\renewcommand\[{\left[}
\renewcommand\]{\right]}

\def\ebq{\end{equation} \begin{equation}}
\renewcommand{\figurename}{Figure.}
\renewcommand{\tablename}{Table.}
\newcommand{\Slash}[1]{{\ooalign{\hfil#1\hfil\crcr\raise.167ex\hbox{/}}}}
\newcommand{\bra}[1]{ \langle {#1} | }
\newcommand{\ket}[1]{ | {#1} \rangle }
\newcommand{\beq}{\begin{equation}}  \newcommand{\eeq}{\end{equation}}
\newcommand{\bef}{\begin{figure}}  \newcommand{\eef}{\end{figure}}
\newcommand{\bec}{\begin{center}}  \newcommand{\eec}{\end{center}}
\newcommand{\non}{\nonumber}  \newcommand{\eqn}[1]{\begin{equation} {#1}\end{equation}}
\newcommand{\laq}[1]{\label{eq:#1}}  
\newcommand{\dd}[1]{{d \o d{#1}}}
\newcommand{\Eq}[1]{Eq.~(\ref{eq:#1})}
\newcommand{\Eqs}[1]{Eqs.~(\ref{eq:#1})}
\newcommand{\eq}[1]{(\ref{eq:#1})}
\newcommand{\Sec}[1]{Sec.\ref{chap:#1}}
\newcommand{\ab}[1]{\left|{#1}\right|}
\newcommand{\vev}[1]{ \left\langle {#1} \right\rangle }
\newcommand{\bs}[1]{ {\boldsymbol {#1}} }
\newcommand{\lac}[1]{\label{chap:#1}}
\newcommand{\SU}[1]{{\rm SU{#1} } }
\newcommand{\SO}[1]{{\rm SO{#1}} }

\def\({\left(}
\def\){\right)}
\def\dt{{d \o dt}}
\def\diag{\mathop{\rm diag}\nolimits}
\def\Spin{\mathop{\rm Spin}}
\def\O{\mathcal{O}}
\def\U{\mathop{\rm U}}
\def\Sp{\mathop{\rm Sp}}
\def\SL{\mathop{\rm SL}}
\def\tr{\mathop{\rm tr}}
\newcommand{\OR}{~{\rm or}~}
\newcommand{\AND}{~{\rm and}~}
\newcommand{\EV}{ {\rm \, eV} }
\newcommand{\KEV}{ {\rm \, keV} }
\newcommand{\MEV}{ {\rm \, MeV} }
\newcommand{\GEV}{ {\rm \, GeV} }
\newcommand{\TEV}{ {\rm \, TeV} }

\def\o{\over}
\def\a{\alpha}
\def\b{\beta}
\def\c{\varepsilon}
\def\d{\delta}
\def\e{\epsilon}
\def\f{\phi}
\def\g{\gamma}
\def\h{\theta}
\def\k{\kappa}
\def\l{\lambda}
\def\m{\mu}
\def\n{\nu}
\def\p{\psi}
\def\q{\partial}
\def\r{\rho}
\def\s{\sigma}
\def\t{\tau}
\def\u{\upsilon}
\def\v{\varphi}
\def\w{\omega}
\def\x{\xi}
\def\y{\eta}
\def\z{\zeta}
\def\D{\Delta}
\def\G{\Gamma}
\def\H{\Theta}
\def\L{\Lambda}
\def\F{\Phi}
\def\P{\Psi}
\def\S{\Sigma}
\def\me{\mathrm e}
\def\ol{\overline}
\def\tl{\tilde}
\def\*{\dagger}

\newcommand{\exclude}[1]{}

\def\bra{\langle}
\def\ket{\rangle}
\def\beq{\begin{equation}}
\def\eeq{\end{equation}}
\newcommand{\C}[1]{\mathcal{#1}}
\def\ov{\overline}

\preprint{TU-1158}

\title{High Energy Sphalerons for Baryogenesis at Low Temperatures}

\author{Joerg Jaeckel}
\affiliation{Institut f\"ur theoretische Physik, Universit\"at Heidelberg,
 Philosophenweg 16, 69120 Heidelberg, Germany}
\author{Wen Yin}
\affiliation{Department of Physics, Tohoku University,  
Sendai, Miyagi 980-8578, Japan}

\begin{abstract}
We discuss baryogenesis in scenarios where the Universe is reheated to temperatures
$\lesssim 100$\,GeV by the decay of long-lived massive particles into energetic SM particles.
Before its thermalization, the center-of-mass energy in collisions between such a particle and a particle from the ambient plasma can be higher than the typical sphaleron mass, even if the temperature of the plasma itself is much lower. 
Optimistic estimates for the high energy enhancement of the sphaleron cross section suggest that successful baryogenesis is possible for reheating temperatures as low as $0.1\text{-}1\,$GeV.
With a simple extension of the SM, 
sufficient baryon production can be achieved even if more pessimistic results for the sphaleron rate are correct.
In both cases this scenario can be probed in collider and cosmic-ray experiments. 
We briefly discuss the possible origin of the required CP violation.

\noindent
\end{abstract}
\maketitle
\flushbottom

{\bf Introduction.--}
Many scenarios of string~\cite{Polchinski:1998rq,Grana:2005jc,Blumenhagen:2006ci,Gorlich:2004qm,Kachru:2003aw} or supersymmetric theories~\cite{Giudice:1998xp, Randall:1998uk,Arkani-Hamed:2004zhs,Giudice:2004tc,Wells:2004di,Ibe:2006de}, (see, {e.g.,}~\,\cite{Ibe:2011aa, Arkani-Hamed:2012fhg,PardoVega:2015eno, Yin:2016shg, Yanagida:2019evh,Yin:2021mls} for discussions {in light} of current data {or in the context of axions~\cite{Banks:2002sd,Visinelli:2009kt,Cicoli:2012aq, 
Higaki:2012ar,Hebecker:2014gka,Angus:2014bia,Cicoli:2015bpq,Nelson:2018via,Cicoli:2018cgu,Acharya:2019pas,Jaeckel:2021gah,Angus:2021jpr,Arias:2021rer,Jeong:2021yol,Frey:2021jyo, Cicoli:2021tzt,Arias:2021rer,Jaeckel:2021ert,Hebecker:2022fcx})} predict relatively low temperatures for the final reheating before the start of standard cosmology. This is due to the existence of intermediate mass particles with Planck suppressed couplings to the standard model (SM) particles. Featuring such small couplings they are often long-lived. Due to the Hubble expansion they then become non-relativistic and eventually dominate the energy density of the Universe. Finally, they decay to reheat the Universe with a relatively low reheating temperature, often below the electroweak scale. 
This is often viewed as a problem for baryogenesis (see, however~\cite{Affleck:1984fy,Elor:2018twp,Alonso-Alvarez:2019fym, Dimopoulos:1987rk,Barbier:2004ez, Babu:2006xc, Grojean:2018fus, Pierce:2019ozl,McKeen:2015cuz, Aitken:2017wie, Asaka:2019ocw, Azatov:2021irb} for examples of mechanisms to achieve the required matter-antimatter asymmetry at potentially quite low temperatures that use additional sources of baryon number violation c.f. Refs.\,\cite{Fukugita:1986hr,Lazarides:1991wu, Asaka:1999yd, Hamaguchi:2001gw}).
The reason is that the sphaleron/instanton effect, which is the only baryon number violating process within the Standard Model, is highly suppressed at temperatures much lower than the sphaleron scale $M_{\rm sph}\sim 7\,$TeV. Nevertheless, we want to argue that even in such a scenario successful baryogenesis may be achieved in a rather minimal setup. 

From the viewpoint of the Sakharov conditions~\cite{Sakharov:1967dj} the late-time decays of heavy particles also provide an opportunity: they are inherently an out-of-equilibrium process.\footnote{See Refs.~\cite{Lazarides:1991wu, Asaka:1999yd, Hamaguchi:2001gw,Barbier:2004ez, Dimopoulos:1987rk, Babu:2006xc,McKeen:2015cuz, Aitken:2017wie, Elor:2018twp,  Grojean:2018fus, Hamada:2018epb, Pierce:2019ozl,Asaka:2019ocw} for baryogenesis using this, too.}
Therefore, let us focus on the short period soon after a long-lived particle decays into energetic SM particles. 
During the first steps of the thermalization of an energetic SM particle, the center-of-mass energy between it and a particle in the (already thermalized) ambient plasma can easily be higher than $M_{\rm sph}$, even when the reheating temperature $\lesssim 100$\,GeV. 
In such energetic collisions, it may be easier to produce suitable sphalerons and therefore fulfill Sakharov's condition of a sufficiently effective baryon number violating process. 

Earlier work in similar directions includes~\cite{Asaka:2003vt}, where the authors considered the possibility that in each decay of the long-lived particles the decay products first heat a small asymmetric region where the plasma temperature is much higher than $100\GEV$. Thus less suppressed symmetric-phase sphalerons can be active in this region. Another option is to use symmetric-phase sphalerons at the early stage of reheating~\cite{Davidson:2000dw} (when the plasma temperature is actually higher), the resulting asymmetry is, however, typically diluted by $(T_R/100\GEV)^5$. In contrast, we will consider the direct sphaleron enhancement by the scattering of an energetic particle from the decay and a plasma particle. 

At present calculations of highly energetic sphaleron rates are still plagued by uncertainties. 
Therefore, we first quantify how much enhancement of the sphaleron effect is needed for successful baryogenesis. The relevant cross sections turn out to be of a size that may allow for tests at the LHC, future colliders, and cosmic-ray experiments.
Turning to concrete calculations of the sphaleron rate we find that more optimistic estimates give values that allow for successful baryogenesis.
If more pessimistic estimates turn out to be correct, we can still obtain successful baryogenesis by considering a simple BSM extension of the gauge group that enhances the gauge interaction and thus the sphaleron rate at high energies.

Finally, turning to the last Sakharov condition -- CP violation --  we briefly discuss some possibilities how this may be realized in the present context in  Appendix~A.

{\bf Brief review of out-of-equilibrium sphalerons.--}
The electroweak instanton/sphaleron is the only source of baryon number violation within the SM. 
As a tunneling process between vacua with different Chern-Simons numbers it is expected that the instanton effect is exponentially suppressed.
A sphaleron, being externally supplied with enough energy, $\gtrsim M_{\rm sph}\sim \pi m_W/\a_w\sim 7\TEV$, 
does not have this suppression since it is a classical saddle-point solution connecting the different vacua. Here, $m_W~(\a_w)$ is the weak boson mass (coupling constant).
Therefore, cosmological models with a reheating temperature $T_R \ll M_{\rm sph}$ are usually argued not to have efficient baryogenesis via sphalerons.

The instanton/sphaleron-induced cross-section for highly energetic quarks has been studied in the context of collider and cosmic ray phenomenology (cf, e.g.,~\cite{Morris:1993wg,Gibbs:1995bt,Ringwald:2002sw,Fodor:2003bn,Brooijmans:2016lfv,Ellis:2016dgb,Ringwald:2018gpv,Papaefstathiou:2019djz}), but also for a better conceptual understanding of the sphaleron itself. 
Unfortunately,  the size of the cross-section for this non-perturbative process is still under debate. 
Let us briefly review the current status.
The instanton effect can be estimated from the t 'Hooft operator~\cite{Callan:1977gz}  ${\cal L}_{\rm eff} \sim e^{-\frac{2\pi}{\a_w} }\frac{Q^9 L^3 }{v^{14}}+h.c.$, 
with $Q$ and $L$ being the left-handed quarks and leptons, and $v\sim 170\GEV$ the Higgs vacuum expectation value (VEV). 
Importantly, \cite{Ringwald:1989ee, Espinosa:1989qn} pointed out that by emitting multiple bosons in 2 to many processes 
$
QQ\to 7 \bar{Q} +3 \bar{L}+n_Z Z +n_WW+n_H H
$
the cross section can be enhanced exponentially in $N=n_Z+n_W+n_H$,
{fitting the picture that sphalerons are more efficient than instantons} at high energy scales since $N$ can be larger. 

The inclusive cross section is often given in terms of the so-called holy grail function, $F$, specifying the exponential suppression,
\beq
\laq{sphacross}
\sigma_{\rm sph}\approx \frac{1}{E_{\rm cm}^2}
\exp{\left(-\frac{4\pi}{\alpha_w} F(E_{\rm cm})\right)},
\eeq
where $E_{\rm cm}$ is the center-of-mass energy.
The holy grail function has been calculated by various authors and using a range of different methods and limits.   
We summarize the different estimations in Fig.\,\ref{fig:0} {as functions of} $\a_w E_{\rm cm}/m_W.$

\begin{figure}[t!]
\includegraphics[width=75mm]{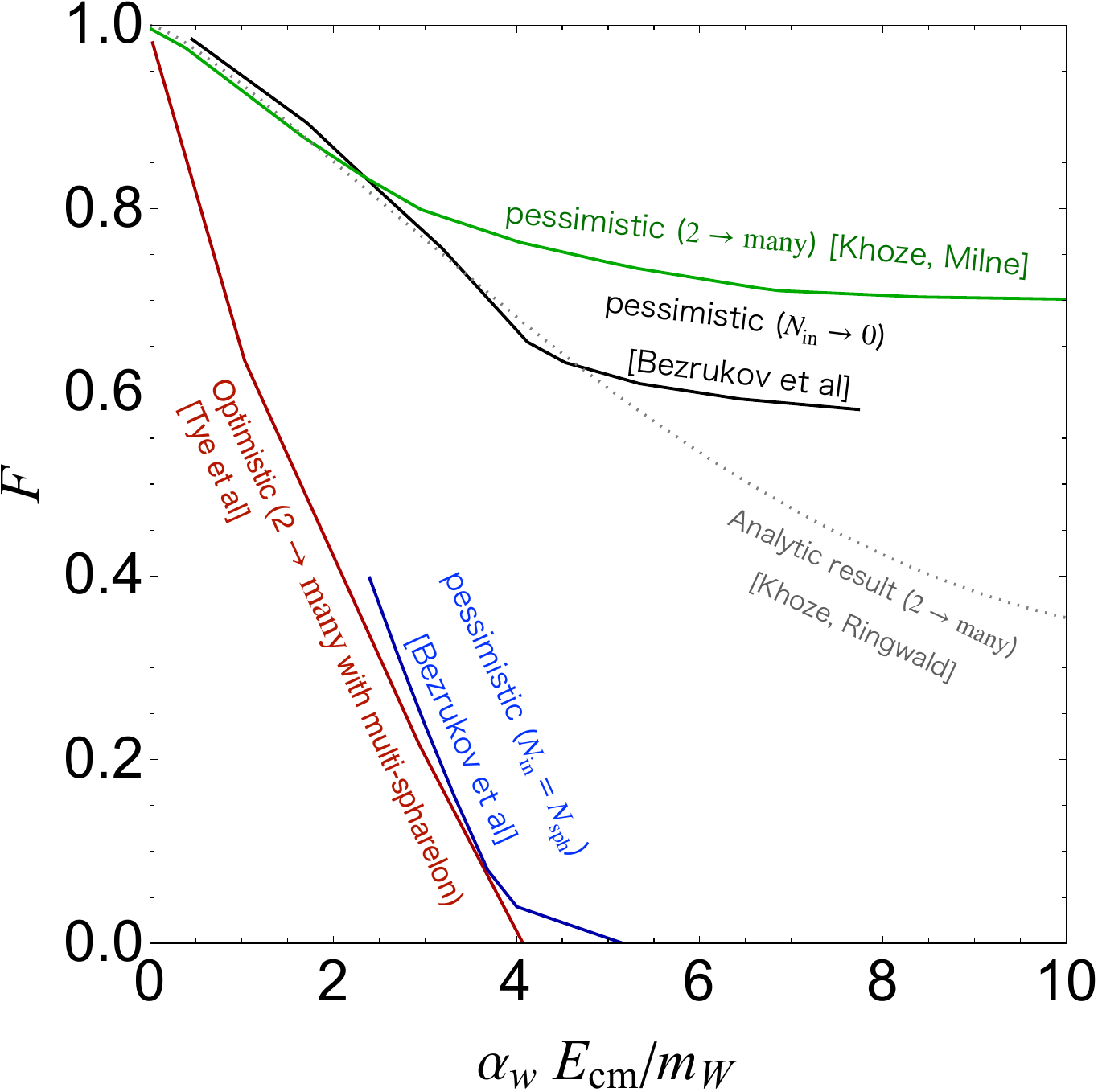}
\caption{Different estimations of the holy grail function depending on $E_{\rm cm}\a_w/m_W$.
The optimistic estimation (red) is from \cite{Tye:2015tva,Tye:2017hfv, Qiu:2018wfb}. The pessimistic ones are from \cite{Bezrukov:2003er, Bezrukov:2003qm} (black) and \cite{Khoze:2020paj} (green). We also show the low energy analytic result~\cite{Khoze:1990bm} (gray dotted) and a result with $N_{\rm in}=N_{\rm sph}=1/\a_w$ translated from \cite{Bezrukov:2003er, Bezrukov:2003qm} (blue). While the latter is in principle a pessimistic estimate, we note that applying it to $\a_w=\O(1)$ and $N_{\rm in}=N_{\rm sph}=2$ might be optimistic (see the discussion below \Eq{gLH}).
}
\label{fig:0}
\end{figure}

{\bf Broken phase sphaleron from reheating.--} For concreteness, we discuss the thermalization process during reheating caused by a modulus decay. 
Our conclusions do not change if we replace the modulus {by the inflaton, a long-lived fermion, e.g. gravitino/Peccei-Quinn fermion, or interaction with macroscopic objects, as, e.g. a bubble wall collision with the ambient plasma~\cite{Azatov:2020nbe, Azatov:2021irb, Baldes:2021vyz} or with} another bubble wall \cite{Katz:2016adq}, as long as they provide energetic daughter SM particles\footnote{In the bubble wall case, especially for the wall-wall collisions, the scenario of \cite{Asaka:2003vt} may also be interesting for the sphaleron process, since the temperature may locally exceed $100\GEV$.} and reheat the Universe.  

Let $\f$ be the long-lived modulus.
A simple case is that $\f$ decays into a pair of energetic SM particles charged under $\SU(2)_L$\footnote{Precisely {speaking, considering} the chirality suppression, the decay of a scalar to SM fermions should be three-body with large $m_\f$. {Alternatively,} the mother particle could also be a vector boson. On the order of magnitude level this does not change our conclusions. }
\beq
\f \to \chi\bar \chi \equiv L \bar{L},~Q\bar{Q}, W\bar W, ZZ, {\rm or/and}~  h h.
\eeq
If the non-relativistic modulus dominates the Universe, the reheating temperature is given by 
$
T_R\equiv \(g_{\star}\frac{\pi^2}{90}\)^{-1/4} \sqrt{M_{\rm pl}\Gamma_{\f}}.
$
Here $M_{\rm pl}\approx 2.4\times 10^{18}\GEV$ is the reduced Planck scale, $g_{\star}$ (and $g_{s\star}$ used below) is the number of relativistic degrees of freedom (for entropy) in the SM {(we use the numbers from~Ref.\,\cite{Husdal:2016haj})}. 
With $\G_\f = m_\f^3/M^2$, $m_\f$ being the mass of the modulus and $M$ the typical scale of the coupling we obtain $T_R \approx 20 \MEV \(\frac{M_{\rm pl}}{M}\) \(\frac{11}{g_{\star}}\)^{1/4}\( \frac{m_\f}{100\TEV}\)^{3/2}$.
Thus, if  $m_\phi \ll 10^8\GEV$, $T_R$ is below the weak scale if $M\sim M_{\rm pl}$.

Let us focus on what happens within a Hubble time around $T=T_R$, where the energy density in the moduli and in radiation is roughly equal. 
Baryons produced much earlier will be 
strongly
diluted~\cite{Davidson:2000dw} and much later 
the sphaleron is inefficient as usual. 
In general, $F$ depends on $E_{\rm cm}$ and the sphaleron contribution during reheating may depend on the behavior of $F$. However, for large enough $E_{\rm cm}$, $F$ is thought to be approximately constant. Thus, we consider a simplified situation where this holds. 
We will briefly revisit this assumption below.

Around the relevant time we can estimate the total number density of highly energetic SM particles, $\chi$, produced in the decay of the moduli, as,
\beq
n_{\chi}\sim 2n_\f \frac{\G_\f}{H} \sim \frac{g_\star \pi^2 T_R^4}{15 m_\f}
\eeq
with $H$ being the Hubble parameter at that time. In the last approximation we have used $\G_\f\sim H$ and 
the definition of the end of reheating,  $n_\f m_\f \sim g_\star \pi^2 T_R^4/30$.

Now,  let us look at the thermalization of the energetic SM particles, $\chi$. 
To this end, we focus on a single $\f$ particle. 
At this time, the Universe is comprised of the other $\f$ particles who will decay later, ambient thermalized plasma from the $\f$ decays at earlier times, 
and the energetic $\chi \bar\chi$ from the $\f$ decay of interest.
Initially $\chi$ carries an energy,
\beq
E_\chi \simeq \frac{m_\phi}{2}.
\eeq
Thermalization proceeds via the gauge interactions.
 The time-scale on which $\chi$ looses an $\O(1)$ fraction of its energy, is $t_{\rm th}$.
 This is estimated ~\cite{Allahverdi:2002pu,Asaka:2003vt, Kurkela:2011ti, Harigaya:2014waa} (see also Refs.\cite{Davidson:2000er,Baier:2000sb}) from the Landau-Pomeranchuk-Migdal effect~\cite{Landau:1953um,Migdal:1956tc} to be,
\beq 
\laq{LPM}
t_{\rm th} \simeq t_{\rm LPM}\simeq 
\(\alpha_\chi ^2 \sqrt{\frac{T_R}{E_\chi} }T_R\)^{-1}.
\eeq
Here, $\alpha_\chi=\alpha_3\,\, \mathrm{or}\,\, \alpha_e$, i.e. the coupling constants for the gluon and the photon, depending whether $\chi$ is a left-handed quark or charged lepton.
The scattering rate (denominator in Eq.~\eq{LPM}) also receives contributions of order $\a_2^3 T_R^3/m_{W,Z}^2 \times m_{W,Z}/E_\chi $ from the W/Z Bremsstrahlung.
If this rate is larger than the denominator on the r.h.s of \eq{LPM}, we should use it in the evaluation of $t_{\rm th}$.
A similar t-channel $2\to 2$ scattering process obtained by exchanging a soft W-boson is particularly important when $\chi$ is a neutrino. In this case, the neutrino is translated into a charged lepton at a time scale~(see e.g. \cite{Jaeckel:2020oet}) $t_{\nu}\approx (\alpha_2^2 T_R^3/m_{W}^2)^{-1}$. After this the resulting charged lepton quickly thermalizes. 
In addition, if the sphaleron reaction is too fast, we need to consider its effect on the thermalization. 
Since \Eq{LPM} will be dominant in most of the parameter region of interest, 
  we will mainly focus on that time-scale in our discussion. In the result of Fig.~\ref{fig:1}, we also include the other contributions.  
The important thing is that during $t_{\rm th}$ we have energetic SM particles, $\chi$, whose collision center-of-mass energy can be much larger than $T_R.$

The most significant process for the baryon number violation is that $\chi$ scatters with the ambient plasma.\footnote{The number density of direct sphaleron-interactions between two highly energetic $\chi$
is further suppressed by $t_{\rm th} H$ and $n_\chi/n_r$ compared to the $\chi$ scattering with the ambient plasma. We expect this to be small unless the higher $E_{\rm cm}$
increases the sphaleron cross section significantly.}
The typical center-of-mass energy for these interactions is $E_{\rm cm}\sim \sqrt{E_\chi T_R}$.  
For each $\chi$, the probability to interact via a sphaleron 
can be estimated as, 
\beq
P_{\rm sph} \simeq  \vev{\s_{\rm sph} n_{r}}\times t_{\rm th}\sim \frac{1}{E_\chi } T_R^2 e^{-\frac{4\pi}{\a_w} F(T_R E_\chi)}\times t_{\rm th}. 
\eeq
Here, $\vev{}$ denotes the thermal average, and $n_{r}$ is the number density of the ambient plasma particles that are charged under the 
$\SU(2)_L$.\footnote{Assuming, again, that $F$ depends only weakly on $E_{\rm cm}$ and the type of incoming particles, we obtain the thermal average of the reaction $\vev{\s_{\rm sph} n_r}\simeq \frac{g\log{\epsilon}}{48 (E_\chi T_R)} T_R^3$ with $g$ being the number of fermion scatterars in the plasma, and $\e$ the IR cutoff in the angular integral. It is roughly $\frac{1}{E_{\chi} T^2_R} T_R^3$ by noting $g \log \e=\O(10).$ } Then we obtain the sphaleron events per unit volume per unit Hubble time as 
\beq
n_{\rm sph}\simeq P_{\rm sph} n_\chi.
\eeq
Normalizing this to the entropy density, $s=\frac{2g_{s\star} \pi^2T_R^3}{45}$, we obtain (using Eq.~\eq{LPM}),
\begin{align}
\frac{n_{\rm sph}}{s}&\sim 3\times 10^{-5} \frac{g_{\star}}{g_{s \star}}e^{-\frac{4\pi}{\alpha_w} F(\sqrt{T_R E_\chi} )}\non \\
~~~&\times \(\frac{10^{7}\GEV}{m_\f}\)^{3/2}\(\frac{T_R}{10\GEV}\)^2 \(\frac{\a_e}{\a_\chi}\)^2.
\end{align}
In the case where the sphaleron interaction dominates the thermalization of $\chi$, it is this rate that determines, $t_{\rm th}^{-1}\sim t^{-1}_{\rm sph}\equiv \vev{\s_{\rm sph} n_r}.$ 
This leads to 
\beq 
\frac{n_{\rm sph}}{s} \simeq  \frac{3g_{\star}}{4g_{s\star}}\frac{T_R}{E_\chi },
\eeq
which is independent of $\s_{\rm sph}$ as long as the reaction is faster than the other thermalization processes.

The baryon to entropy ratio should satisfy \cite{Aghanim:2018eyx}
\beq
\laq{nB}
\frac{\D n_B}{s}\simeq \e_{\rm CPV} \frac{n_{\rm sph}}{s}\simeq 8.7\times10^{-11}\,
\eeq
with $\e_{\rm CPV}$ being a model-dependent parameter representing the amount of the CP violation, which will be discussed in Appendix~A. 
Effectively it is the number of baryons produced per sphaleron interaction. Usually, we expect $\e_{\rm CPV}<1.$ In Fig.~\ref{fig:1} we show the required sphaleron cross section (red dashed) and values for the holy grail function (black) that allow for a sufficient baryon asymmetry in the
$T_R-E_\chi (\simeq m_\f/2)$ plane (for $\e_{\rm CPV}=10^{-2}$). 
We have taken into account the various thermalization processes assuming that $\chi$ is a charged lepton, and using $t_{\rm th}^{-1}= t^{-1}_{\rm LPM}+t_{\rm sph}^{-1}$.
The lower shaded region corresponding to $\sqrt{T_R E_\chi} \lesssim \TEV$ may be inconsistent with 
 the collider search~\cite{Ringwald:2018gpv, Papaefstathiou:2019djz}. 
In the blue region, the sphaleron interaction dominantly thermalizes $\chi$  and not enough baryons are generated. {Fig.~\ref{fig:1-2} in Appendix~B depicts} the case where $\chi$ is a neutrino where the soft W boson exchange can be important.

\begin{figure}[t!]
\includegraphics[width=60mm]{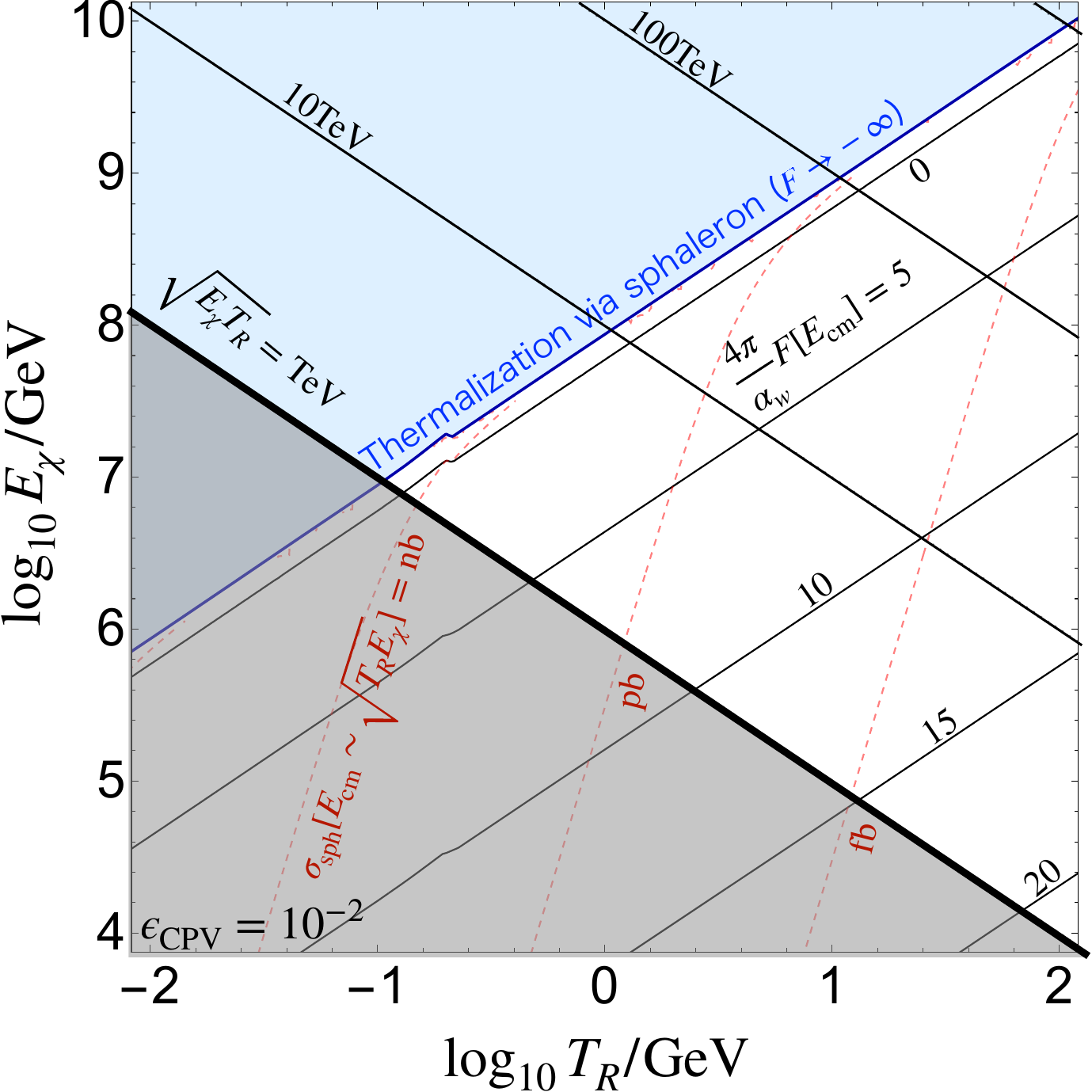}
\caption{Contours for the required sphaleron cross-section (red dashed lines) and holy grail function $4\pi F /\a_w$ (thinner black lines) at the center-of-mass energy $E_{\rm cm}\sim \sqrt{T_R E_\chi}$ for 
successful baryogenesis with $\epsilon_{\rm CPV}=10^{-2}$ as a function of $T_R$ and $E_\chi (\approx m_\f/2)$. Here, we assume $\chi$ is a charged lepton. 
We show the contours of $E_{\rm cm}\sim \sqrt{E_\chi T_R}=1,10, 100\TEV$ in black bold lines. The shaded region below the former may be already excluded by collider searches. 
Above the blue solid line, the sphaleron dominates the thermalization and the baryon asymmetry is too small.
This analysis is only valid when $E_{\rm cm}$ is in the regime that $F$ is almost a constant. }
\label{fig:1}
\end{figure}

Exploring the parameter space (cf. also Figs.~\ref{fig:1},~\ref{fig:1-2}) we find that successful baryogenesis is possible even for temperatures around $\sim 100$~MeV, if the sphaleron cross sections are large enough.
If we adopt the most optimistic estimation of Fig.~\ref{fig:0}, $F\approx0$ for $E_{\rm cm}\gtrsim 9\TEV$. In this case, our scenario allows for $T_R\lesssim 100\GEV$ with $\e_{\rm CPV}\gtrsim 10^{-6}$.
As mentioned before, realistically $F$ is not a simple constant. 
In the optimistic scenario of Fig.~\ref{fig:0}, $F$ changes rapidly when $E_{\rm cm}<9\TEV$.  If $\sqrt{T_R E_\chi} \lesssim 9\TEV$, the reaction during reheating with the dilute plasma temperature $T>T_R$ is actually more important despite the later dilution. It is dominant at $T \sim T_*$  satisfying $\sqrt{T_* \cdot E_{\chi}}\sim 9\TEV$ if the reheating lasts long enough. For this case we obtain $\frac{\Delta n_B}{s}\simeq \e_{\rm CPV}\(\frac{T_R}{T_*}\)^5\left.\frac{n_{\rm sph}}{s}\right|_{T_R\to T_*}$ if $T_R< T_*$.\footnote{When $T_*\gtrsim 100\GEV$, the usual sphaleron in the thermal plasma is active and we also need to consider the effect of~\cite{Davidson:2000dw}.} From this we may obtain additional allowed parameter space.

However, if we adopt the pessimistic estimations in Fig.~\ref{fig:0}, we get $4\pi/\a_w F= \O(100)$, and the scenario is difficult in the minimal setup. Therefore, let us briefly consider BSM effects that may sufficiently enhance the sphaleron such that successful baryogenesis is also possible in the pessimistic case. 

{\bf  BSM with faster sphalerons.--}
Let us consider a simple renormalizable model to enhance the sphaleron rate.
The key idea is to enhance the gauge coupling for the SM fermion at a high energy scale. 
One possibility is $\SU(2)_1\times\SU(2)_2\to \SU(2)_L$ broken by a bi-fundamental Higgs field. (See also models with a bi-fundamental Higgs for 
GUT breaking~\cite{Yanagida:1994vq,Izawa:1997he} and generating axion/ALP potentials~\cite{Agrawal:2017ksf,Csaki:2019vte,Gherghetta:2020ofz,Takahashi:2021tff,Takahashi:2021bti}.)
The Lagrangian is given as
\beq
\laq{BSM}
{\cal L} =- \sum_{i}\frac{1}{4 g_1^2}F_{1}F_1+{\cal L}_{H_b}+{\cal L}_{\rm SM}(A_2)\,,
\eeq
with ${\cal L}_{H_b}$ being the Lagrangian for a bi-fundamental Higgs field, which transforms as $H_b \to U_1^\* H_b U_2$.
 $F_i$ ($U_i$) is the field strength (gauge transformation) for the gauge field $A_i$ of $\SU(2)_i$. We omit the Lorentz indices. 
$A_1$ ($A_2$) does not (does) appear in ${\cal L}_{\rm SM}[A_2]$ which is the SM Lagrangian with the usual $SU(2)_L$ fields and coupling, $A_L, g_L$, replaced by $A_2, g_2$. 

It is straightforward to find a renormalizable potential such that 
$\vev{H_{b}}=v_b\delta_{ij},$.
This spontaneously breaks $\SU(2)_1 \times \SU(2)_2\to \SU(2)_L$, absorbing $2\times (2^2-1)-(2^2-1)$ would be Nambu-Goldstone bosons.  
For simplicity, we further restrict the field space with the condition,
\beq 
\laq{reality}
H_b= \sigma_2 H_b^* \sigma_2
\eeq 
such that the number of real degrees of freedom is four, three of which are eaten by the gauge bosons (see Appendix~C for a brief discussion of the situation where we do not impose this constraint).

There exists a heavy ($A_{H}$) with mass $M^2_H=v^2_b(g^2_{1}+g^{2}_{2}$) and an up to this point massless ($A_L$) gauge boson,
\beq
A_L\equiv \cos\h A_1 +\sin\h A_2,\quad A_H\equiv \sin\h A_1 -\cos\h A_2,
\eeq  
with $\sin \theta= \frac{g_1}{\sqrt{g_1^2+g_2^2}},\,\,\cos \theta= \frac{g_2}{\sqrt{g_1^2+g_2^2}}$.
The light $A_L$ corresponds to the usual weak bosons and couples to to the SM fermions with strength, 
\beq
\laq{gLH}
g_L= \sin\h g_2 \quad g_H=\cos\h g_2.
\eeq

If we have $g_2\gg g_1,$ we get $g_L\simeq g_1$, $A_L\simeq A_1$, as well as $A_H\simeq A_2$ and $g_H\simeq g_2$.
Thus we can have a strong coupling of the SM fermions/Higgs to $A_2$.
To evade collider bounds we need $M_H\gtrsim \O(1)\TEV$. 

Importantly,
when $E_{\rm cm}\gg M_H,$ the strongly coupled $\SU(2)_2$ also contributes to the sphaleron rate, significantly enhancing it. 
For instance, the suppression factor $4\pi/\a_2 \sim 10$ in \Eq{sphacross} 
for $\a_2\sim 1$. Thus even we take $F=1$ the required amount of $4\pi F/\a_2 $ for baryogenesis is achieved.   
In this case, one can have an efficient baryon number violation with $T_R\gtrsim 100\GEV (10\GEV,1\GEV)$ and $m_\f=\TEV$-PeV with $\e_{\rm CPV}=10^{-6}(10^{-3},1)$.

Moreover, for large $\alpha_2$ we may be closer to the {situation $2\sim N_{\rm in}\sim N_{\rm out}\sim 1/\alpha_{2}$} that gives the blue curve in Fig.~\ref{fig:0} that corresponds to a nearly unsuppressed rate\footnote{To have a similar setup, we may consider $\chi =W', H_b, \OR H$.} with $F\approx 0$.\footnote{For smaller values of the gauge coupling one would need multi-particle initial states which are highly suppressed by phase space.}

If the result given in \cite{Bezrukov:2003er, Bezrukov:2003qm} is applicable to $\a_2\sim 1$ with fermions, we {could have successful baryogenesis with $T_R\gtrsim 10\GEV, (100\MEV, 10\MEV)$ with $m_\f=\TEV-$EeV and} $\epsilon_{\rm CPV} =10^{-6}(10^{-3},1)$.

Another efficient, but not renormalizable, model to enhance the sphaleron rate may be a large volume extra-dimension where the $\SU(2)_L$ gauge boson lives in the bulk. 
In this case, the small instanton of the $\SU(3)$ can be significantly enhanced in the context of the heavy QCD axion~\cite{Poppitz:2002ac,Gherghetta:2020keg}. However, the CP-violating higher dimensional terms should be somehow made small to solve the strong CP problem~\cite{Kitano:2021fdl,Demirtas:2021gsq}. In the context of our $\SU(2)$ sphalerons, the CP-violating nature is good news since it naturally satisfies one of the conditions for baryogenesis.

{\bf To conclude.-}
Depending on whether more optimistic or more pessimistic estimates for the sphaleron rate at high energies are correct, successful baryogenesis via sphalerons induced by scattering energetic decay products on the ambient plasma may be achievable for reheating temperatures as low as $\sim 100$~MeV with or without a BSM modification.
This scenario can be probed in collider or cosmic-ray experiments by searching for the sphaleron reaction.
In the BSM models, we can also search for heavy gauge bosons that couple to SM particles.
In addition the $\f$ decays may also produce weakly-coupled out-of-equilibrium BSM particles, e.g. right-handed neutrinos~\cite{Jaeckel:2020oet}, WIMP-like particles~\cite{Jaeckel:2020oet} or axion-like particles~\cite{Cicoli:2012aq,Higaki:2012ar,Angus:2013sua, Conlon:2013isa, Hebecker:2014gka, Evoli:2016zhj, Armengaud:2019uso, Acharya:2019pas, Dror:2021nyr,Jaeckel:2021gah,Jaeckel:2021ert,Hebecker:2022fcx},  that reach Earth. Our scenario could then also be probed by measuring those ``primordial cosmic-rays".  

That said, a further understanding of the sphaleron interaction with multiple incoming particles in the regime of large coupling and including fermions is quite important for clarifying the allowed parameter region of this simple scenario.

\noindent\\[0.1cm]
{\bf{Acknowledgments.--}}
WY was supported by JSPS KAKENHI Grant Number 20H05851, 21K20364, 22K14029, and 22H01215. JJ is happy to be a member of the EU supported ITN HIDDEN (No 860881).

\appendix

\section{CP violation}
So far we have simply parametrized the necessary CP violation by employing a free parameter $\e_{\rm CPV}$ in the sphaleron rate. Let us now list some exemplary scenarios how such a CP violation could arise. In fact, there are various viable models for  CP violation e.g.~\cite{Dimopoulos:1987rk,Akhmedov:1998qx,Barbier:2004ez, Babu:2006xc,Asaka:2005pn, Grojean:2018fus, Pierce:2019ozl,McKeen:2015cuz, Aitken:2017wie, Elor:2018twp,Asaka:2019ocw, Azatov:2021irb}.  In principle, these should also work in our scenario of sphalerons being triggered in the broken phase.
\\
{\bf Baryogenesis via CP-violating sphaleron reactions} We can directly produce the baryon asymmetry by including the CP violation during the thermalization. 
For instance, we can slightly extend \Eq{BSM} without imposing the restriction of the field space. 
Note that we can have the couplings $(\lambda' |H|^2+A'+\lambda''|H_b|^2) \det(H_b^2)+h.c.$ with $H$ being the Higgs doublet that gives masses to the SM fermions.  In this case, the couplings $\l',A',\l''$ generically involve CP phases that cannot be removed by the redefinition of the $H_b$ phase. (They {are absent if we impose} the restriction.) 
Using na\"{i}ve dimensional analysis we expect very roughly that $\e_{\rm CPV}\sim \max{[|\Im[A' \l'^*]/M_{\rm sph}|, |\Im[A' \l''^*]/M_{\rm sph}|, |\Im[\l' \l''^*]|]}.$

With the field space restriction a possibility is to
 introduce another heavy Higgs doublet, $\tl{H}$, whose VEV is small. If it transforms via $(2, -1/2)$ under $\SU(2)_1\AND \U(1)_Y$, 
the Higgs potential can have {a contribution} $V\supset A \tl{H}^* \cdot H_b \cdot H + \tl\l (\tl H^*)^2  H^2$. By noting that the phase of $H_b$ is fixed by the reality condition \eq{reality}, the phase $\arg{\tl \l A^*}$ cannot be removed by field-redefinitions.  
Then very roughly we get $\e_{\rm CPV} \sim \frac{|\Im[\tl \l A^*]|}{M_{\rm sph}}$. 

Interestingly, in this setup the flavor structure of the SM may be partially explained by noting that some families may be charged under $\SU(2)_1$ while some are charged under $\SU(2)_2$.
For instance, let us consider that only the 3rd generation fermions are charged under $\SU(2)_2$ and thus obtain mass via the VEV of $H$.  Then the first two generation masses are naturally suppressed by the mixing between  $\tl H$ and $H$. The flavor structure is similar to the two Higgs doublet model specifying the top quark in Ref.~\cite{Das:1995df}. The only difference is that 
the flavor-dependent discrete symmetry in the reference is replaced by the gauge symmetry in our case. Therefore the flavor structure is quite similar: i.e. we have suppressed flavor-changing neutral currents among the first two generation quarks as well as an
electron electric dipole moment (EDM) which is generated at the 2-loop level~\cite{Das:1995df}. In particular, with a BSM mass scale around  $1-10\TEV$ the EDM may be probed in the not too distant future. 
If the sphaleron is enhanced by a large $\a_2$, there is a small instanton contribution in the low energy physics. In this case we have an effective coupling $\frac{Q^3L}{v_b^2} e^{-2\pi/\a_2[v_b]}$, for the 3rd generation fermions.
Via CKM mixing $V_{\scriptscriptstyle{C}}$, we then obtain an operator that involves only first generation quarks, e.g. $(V_{\scriptscriptstyle{C}})_{t,d}(V_{\scriptscriptstyle{C}})_{t,d}(V_{\scriptscriptstyle{C}})_{t,b} \e_b (V^{*}_{\scriptscriptstyle{C}})_{u,b} \frac{d_L d_L u_L \nu_\tau }{v_b^2} e^{-2\pi/\a_2[v_b]} $ with $\e_b= \O(y_b^2/16\pi^2)$ being the order of {magnitude estimate of} the RG effect and $y_b$ the bottom quark Yukawa coupling.
The effective suppression scale for this higher dimensional operator is around the GUT scale {for} $\a_2\sim 0.2 \AND v_b\sim 10\TEV$. Thus proton decay may be a useful tool to probe this scenario. On the other hand, it is already excluded for $\a_2\sim 1$. 
We can also use different charge assignments. For example if the first two generations are charged under $\SU(2)_2$ the proton does not decay. This is because then the operator generated by the instanton is $\frac{Q^6L^2}{v_b^8} e^{-2\pi/\a_2[v_b]}$, which preserves a $Z_2$ baryon number symmetry, which stabilizes the proton. This model can also explain the small masses of the first two generations if $\tl H$ is the SM-like Higgs field. 
\\
{\bf Leptogenesis} One can also first produce a lepton asymmetry from {the} $\f$ decays. Later this is transferred into the baryon asymmetry by the broken phase sphalerons discussed in this Letter. This is the famous idea of leptogenesis in the symmetric phase~\cite{Fukugita:1986hr,Lazarides:1991wu, Asaka:1999yd, Hamaguchi:2001gw}. One can easily produce the lepton asymmetry from lepton flavor oscillations~\cite{Akhmedov:1998qx,Asaka:2005pn,Hamada:2018epb,Eijima:2019hey}. Alternatively, we can assume that $\f$ carries lepton number and employ the Affleck-Dine mechanism~\cite{Affleck:1984fy,Dine:1995kz}. {The SM leptons are then obtained from a coupling $\f HLHL$}. In any case, we need a much larger neutrino/lepton asymmetry $\sim \e_{\rm CPV}$ than the baryon asymmetry because only a small fraction can be transferred by the sphalerons induced by highly energetic particles (cf.~\Eq{nB}). This is in contrast to the conventional symmetric phase leptogenesis which predicts that the two asymmetries are comparable.  The predicted large neutrino asymmetry may be probed, e.g., in future CMB experiments~\cite{Kinney:1999pd,Bonilla:2018nau}.
\\
{\bf Spontaneous baryogenesis (may not work)}
We may also consider a coupling of $\frac{\a_w}{2\pi}\frac{\O[t]}{\L^d} {\rm tr}W\tl W$ with $\O$ being a time-dependent operator made up of single or several fields, $1/\L^d$ the higher dimensional coupling, and $d$ the dimension of $\O$.
Then in the single sphaleron transition, we get the amplitude 
 ${\cal M} e^{\pm i \(\frac{\O(t+\d t) }{\L^d}\)}\simeq  {\cal M} e^{\pm i(\frac{\O(t)+\dot\O \d t}{\L^d})}$ for particle/anti-particle process with $\d t$ being a small change of time and expanding $\O$ in time up to linear order. This amplitude time-dependence feeds into the energy momentum conservation and thus the cross section $\s_{\rm sph}[E_{\rm cm}]\to \s_{\rm sph}[E_{\rm cm}\pm \frac{\dot\O}{\L^d}]$ for the particle/antiparticle transition. In this way we obtain a CP violating effect $\e_{\rm CPV}\simeq \frac{\dot\O}{\L^d} \frac{\partial}{\partial E_{\rm cm}} \log\s_{\rm sph}$.
A rough estimate tells us $\e_{\rm CPV} \sim \dot \O/(\L^d E_{\rm cm}).$
This is similar to the spontaneous baryogenesis picture~\cite{Cohen:1987vi} (cf. also~\cite{DeFelice:2002ir,Chiba:2003vp,Takahashi:2003db,DeFelice:2004uv,Kusenko:2014uta,Ibe:2015nfa,Takahashi:2015waa,Takahashi:2015ula,Jeong:2018jqe,Bae:2018mlv,Abel:2018fqg,Co:2019wyp,Domcke:2019mnd,Domcke:2020kcp,Co:2020xlh,Im:2021xoy} for a variety of interesting implementations) in which the baryon asymmetry is proportional to the chemical potential $\mu=\frac{\dot{\O}}{\L^d}.$
However, this also requires $\dot{\O} \sim E_{\rm cm} \L^d $ in order to have efficient CP violation. Assuming $\L \gtrsim E_{\rm cm}$, this implies that there is a large kinetic energy density contribution from the operator $\gtrsim E_{\rm cm}^4$ (at least for $d=1$). This is inconsistent with the assumption in our setup that the Universe has an energy density $T_R^4 \ll E_{\rm cm}^4$. Thus the spontaneous baryogenesis scenario may not work. \\[0.1cm]

\section{Sphalerons from neutrino scattering}
\begin{figure*}[t!]
\includegraphics[width=50mm]{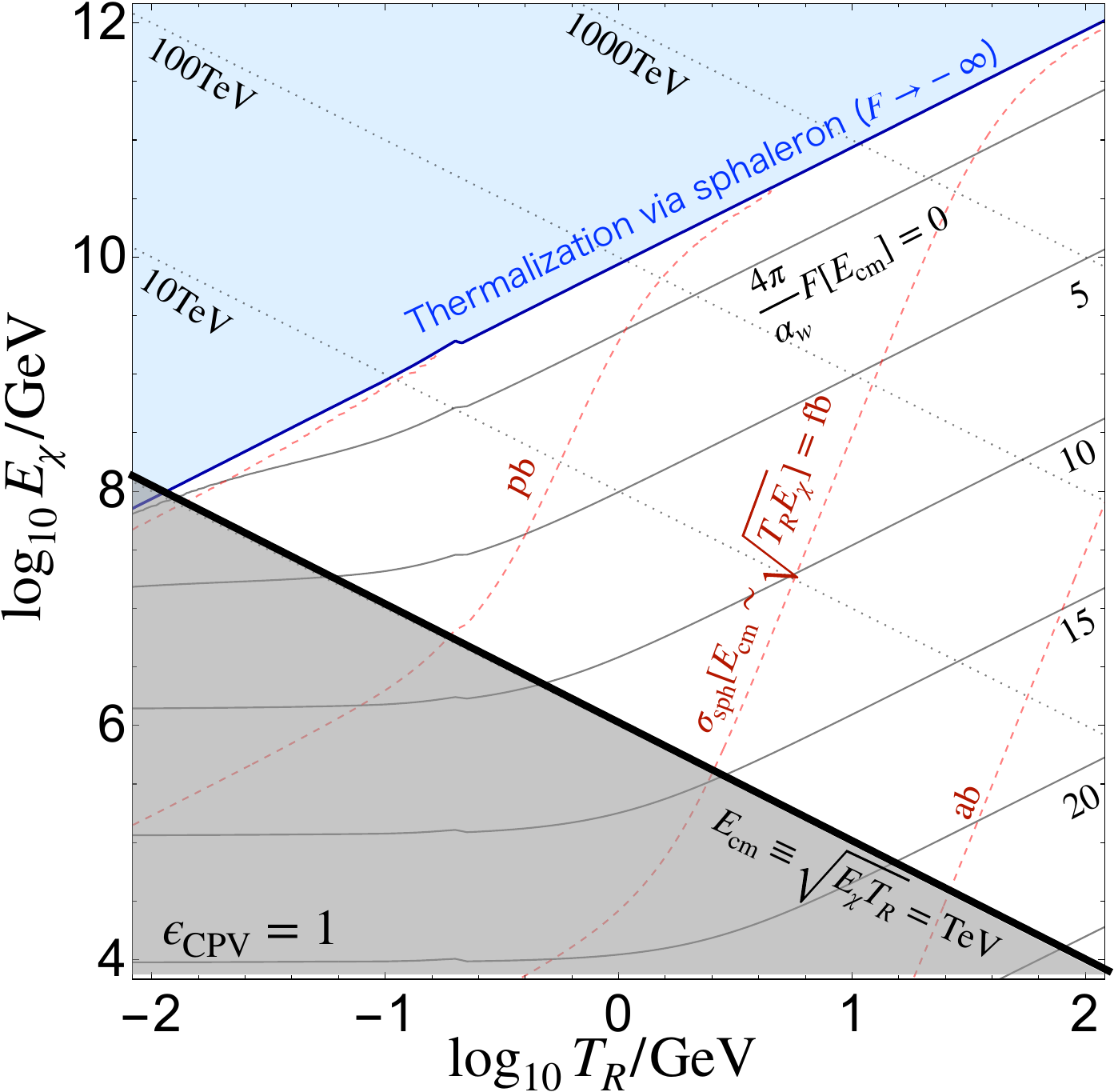}
\includegraphics[width=50mm]{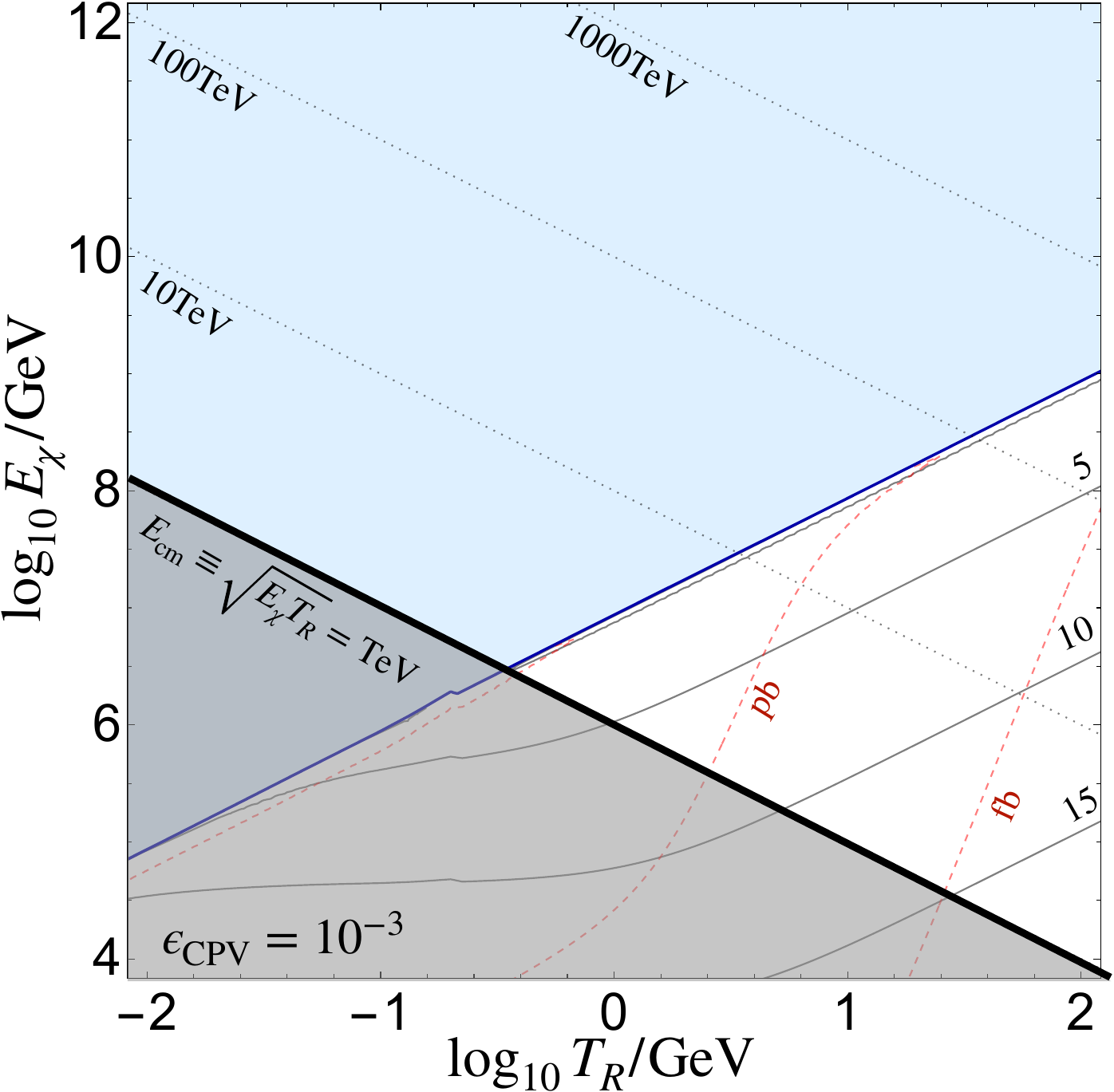}
\includegraphics[width=50mm]{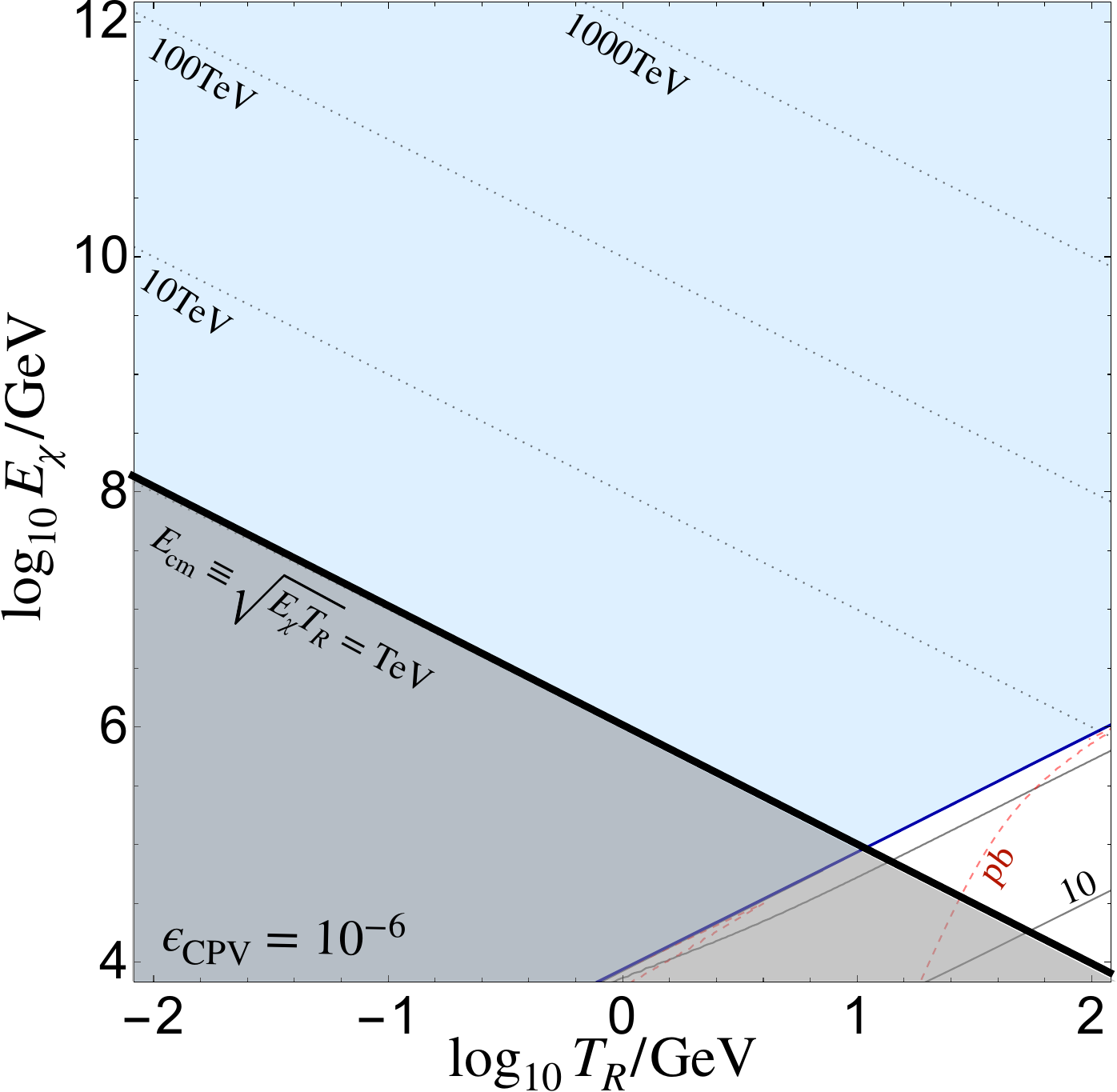}
\caption{Same as Fig. \ref{fig:1} with $\chi$ being a neutrino with $\e_{\rm CPV}=1,10^{-3},10^{-6}$ from left to right.}
\label{fig:1-2}
\end{figure*}
In the main text we have shown results for the situation where the moduli decay into charged leptons. Another interesting situation arises if the decay is into neutrinos.
This is shown in Fig.~\ref{fig:1-2}, where $\e_{\rm CPV}=1$ (top panel), $10^{-3}$ (middle panel), $10^{-6}$ (bottom panel). Here, we 
use $t_{\rm th}^{-1}= (t_{\nu}+t_{\rm LPM})^{-1}+t_{\rm sph}^{-1}$ because only after a soft W-boson exchange, the photon interaction becomes active.
The contour is flat in the low temperature regime because the thermalization time-scale for a neutrino is determined by the W boson emission, $t_{\nu}$. This gives a longer time for the sphaleron reaction to work.
\\

\section{Vacuum structure of the bifundamental Higgs}
Here we discuss the vacuum structure by stabilizing the potential for the bifundamental $\SU(2)_1\times \SU(2)_2$ Higgs.  Let us start by commenting on what happens if we do not impose the restriction \eq{reality}. 
From gauge invariance, the potential is then given  by 
\begin{align}
V= &-\mu_H^2 \tr[H^\*_b \cdot H_b] + \lambda_b (\tr[H^\*_b \cdot H_b])^2 \non \\
&+ \lambda_{\rm tr} \tr[H^\*_b \cdot H_b \cdot H^\*_b \cdot H_b ] \non \\
&+ A' \det(H_b)+\lambda'' (\det H_b) |H_b|^2+h.c. \laq{pottot}
\end{align} 
Here $\lambda_b,\lambda_{\tr}, \lambda''$ denote the dimensionless parameters and $\mu_H, A'$ are the dimensionful parameters.
The first line is invariant under an $\SO(8)$ symmetry. Here, $\lambda''$ and $A'$ may be complex.  
The second line breaks $\SO(8)$ explicitly down to $\U(1)\times \SU(2)_1\times \SU(2)_2$. Finally, the third line breaks the $\U(1)$ explicitly. Therefore, the would-be  Nambu-Goldstone (NG) bosons from the symmetry breaking of $\SO(8)\to \SU(2)_1\times \SU(2)_2$ obtain masses via the explicit breakings. In addition, the third line also contributes to the CP violation, as discussed above.
In general, we also have the potential terms $\d V = \l''' \det[H_b]^2 
+\l'''' \det{H_b}\det{H^\*_b}+\lambda_P |H|^2 \tr[H_b^\* H_b] + \lambda_P' H^\* \cdot H_b^\* \cdot H_b  \cdot H + \lambda' \det{[H_b]} |H|^2+h.c. $ with additional dimensionless couplings.  
Including them does not significantly change the discussion as long as the additional couplings are subdominant.

To check the spectrum, we expand 
\beq H_b= {\bf 1} T_0+ \sum_i \sigma_i T_i \eeq
by using the Pauli matrices. Here $T_{0,1,2,3}$ are the complex fields. We then find
\begin{eqnarray}
H^\dagger_b H_b&=&{\bf 1}(\ab{T_0}^2+ \sum_i|T_i|^2)
\\\nonumber
&&\qquad\qquad+  \sum_i\sigma_i( T_i^*T_0+T_0^* T_i+i \epsilon_{ijk}T_j^* T_k).
\end{eqnarray}

 If we take $\mu_H^2>0, \lambda_b>0$ and neglect the other terms the potential becomes \beq V_1=-2\mu_H^2(|T_0|^2+\sum_i |T_i|^2)+4\lambda_b(|T_0|^2+\sum_i |T_i|^2)^2.\eeq
Thus
$ \vev{H_b}= \vev{T_0}{\bf 1}=  \frac{1}{2}\sqrt{\frac{\m_H^2}{\lambda_b}}{\bf 1}
$ without loss of generality. At this point,  we have $7$ NG modes. 

Next let us introduce $\lambda_{\tr}\neq 0$, but $|\lambda_{\tr}|\lesssim \l_b/2.$
Then we can see that the potential features an additional term
\begin{eqnarray}
\laq{V2}
V_2&=& 2\lambda_{\rm tr}\bigg( (|T_0|^2+\sum_i |T_i|^2)^2
\\\nonumber
&&\qquad\quad+ \sum_i( T_i^*T_0+T_0^* T_i+i \epsilon_{ijk}T_j^* T_k)^2\bigg).
\end{eqnarray}
Now from the first term the VEV is shifted to 
$\vev{T_0}= \sqrt{\mu_H^2/(\lambda_b+\lambda_{\rm tr}/2)}/2.$ We note that the second term in \Eq{V2} provides a mass term for the CP-even triplet
\beq 
V_2\supset 2\lambda_{\rm tr}|\vev{T_0}|^2\sum_i (T_i^*+T_i)^2 . 
\eeq 
The CP-odd triplets are the would-be NG bosons that are eaten by the heavy gauge bosons in the $\SU(2)_1\times \SU(2)_2\to \SU(2)_L$ breaking.
To {achieve that the remnant $\SU(2)$ remains unbroken,} i.e. $T_i^*+T_i$ does not have a VEV, we need $\lambda_{\rm tr}>0.$ We also note that given $\lambda_{\rm tr}=\O(1)$ the triplet is as heavy as $\vev{T_0}=v_b.$

The determinant terms not only give a VEV shift and the mass shift to the CP-even triplet Higgs but it also gives the mass to the neutral NG boson associated with the $\U(1)$ symmetry, under which, $H_b\to e^{i\alpha}H_b.$ 
We get the potential for the axion $a$ from the term as
\beq 
(A' v_b^2+ \l'' v_b^4) e^{i a/{v_b}}+h.c.
\eeq 
In summary, having spontaneous symmetry breaking by our bifundamental Higgs and a potential that consists only of the first row of \Eq{pottot}, we obtain 7 NG modes. They are a neutral ``axion" and 2 triplet Higgs fields under $\SU(2)_L$. The CP-odd triplet is eaten by the gauge boson. 
The second row of \Eq{pottot} provides mass for the CP-even triplet Higgs. The third row finally provides masses to the CP-even triplet Higgs and the ``axion." 
Our discussion in the main part corresponds to the case that the triplet Higgs and the ``axion" are heavy.
If they are lighter they may lead to an interesting phenomenology in accelerator experiments.

\bibliography{references}

\end{document}